\documentclass{PoS}

\usepackage{url}
\usepackage{epsfig}
\usepackage{subfigure}
\newcommand{\beq}{\begin{equation}}
\newcommand{\eeq}{\end{equation}}

\definecolor{literal}{rgb}{0.58,0,0.82}

\title{
Public repository with Monte Carlo simulations for high-energy particle collision experiments
}

\ShortTitle{HepSim Monte Carlo repository}

\author{\speaker{Sergei Chekanov}\\
HEP Division, Argonne National Laboratory,
9700 S. Cass Avenue,
Argonne, IL 60439, USA \\ 
E-mail: \email{chekanov@anl.gov}}

\abstract{ 
Planning high-energy collision experiments for the next few decades requires 
extensive Monte Carlo simulations in order to accomplish physics goals of these experiments.
Such simulations are essential for understanding fundamental physics processes, as well as for setting up the
detector parameters that help establish R\&D projects required over the next few decades.
This paper describes a public repository with
Monte Carlo event samples before and after detector-response simulation. 
The goal of this repository 
is to facilitate the accomplishment of many goals in planning
a next generation of particle experiments.
}

\FullConference{38th International Conference on High Energy Physics\\
                3-10 August 2016\\
                Chicago, USA}

\begin{document}

\section{Introduction}

The High-Luminosity Large Hadron Collider (HL-LHC) can answer many essential questions of high-energy physics (HEP) 
over the next few decades.
Other possible HEP projects   
are the High-Energy LHC (HE-LHC), Large Hadron electron Collider (LHeC), International Linear Collider (ILC), Circular Electron Positron Collider (CEPC) in China,
Super Proton-Proton Collider (SPPC), 
and the Future Circular Collider at CERN (FCC-ee, FCC-ep, and FCC-hh).
Although the technical plans of these projects should still be shaped, the fact that they are on the map of 
possible projects 
shows that HEP is a vibrant field with promising future. 
To secure this future, exploratory physics studies and R\&D projects focused on detector designs should be established.  
Such studies require Monte Carlo collision events  
after fast or full (Geant4) simulation of detector response.
Analysis of these Monte Carlo simulations after a realistic event reconstruction  
is crucial for understanding physics processes, exploring technology choices, detector parameters, 
and for seeding new hardware R\&D projects.
 
Due to the diversity of future HEP experiments, 
leveraging Monte Carlo simulations  created by a large community of researchers is 
an effective way to advance physics studies, detector designs and  analysis techniques.  
The HepSim  project \cite{Chekanov:2014fga} described in this paper 
is a recent attempt to create 
an open-access data catalogue with Monte Carlo predictions for the HEP community. 
This paper describes some of its features.

\section{HepSim repository}

The goal of the HepSim repository is to provide the HEP community with events 
from Monte Carlo event generators, as well as with events 
after full and fast detector simulations. 
This data repository was started at ANL during the US long-term planning 
study of the American Physical Society's Division of Particles and Fields (Snowmass 2013). 
Currently, HepSim hosts  about 220 Monte Carlo event samples,  totaling  
1.6 million events. A single data sample contains a set of files from Monte Carlo event generators
with fully documented information on Monte Carlo settings. 
The data created by event generators will further be called EVGEN events. 
A fraction of EVGEN events have been processed
through fast and full detector simulations using several detector designs.

HepSim repository uses the computational capabilities of the Chicago area, 
such as Open-science grid Connect (OSG-CI, U.Chicago/ANL), 
BlueGene/Q supercomputer of the Argonne Leadership Computing Facility,
Argonne's Laboratory Computing Resource Center (LCRC),
and ATLAS support center (HEP/ANL).
Several data storages are provided by the HEP/ANL, Univ. of Chicago (OSG-CI) and
the Science Gateways at NERSC.

\begin{figure}
\centering
\includegraphics[width=.7\textwidth]{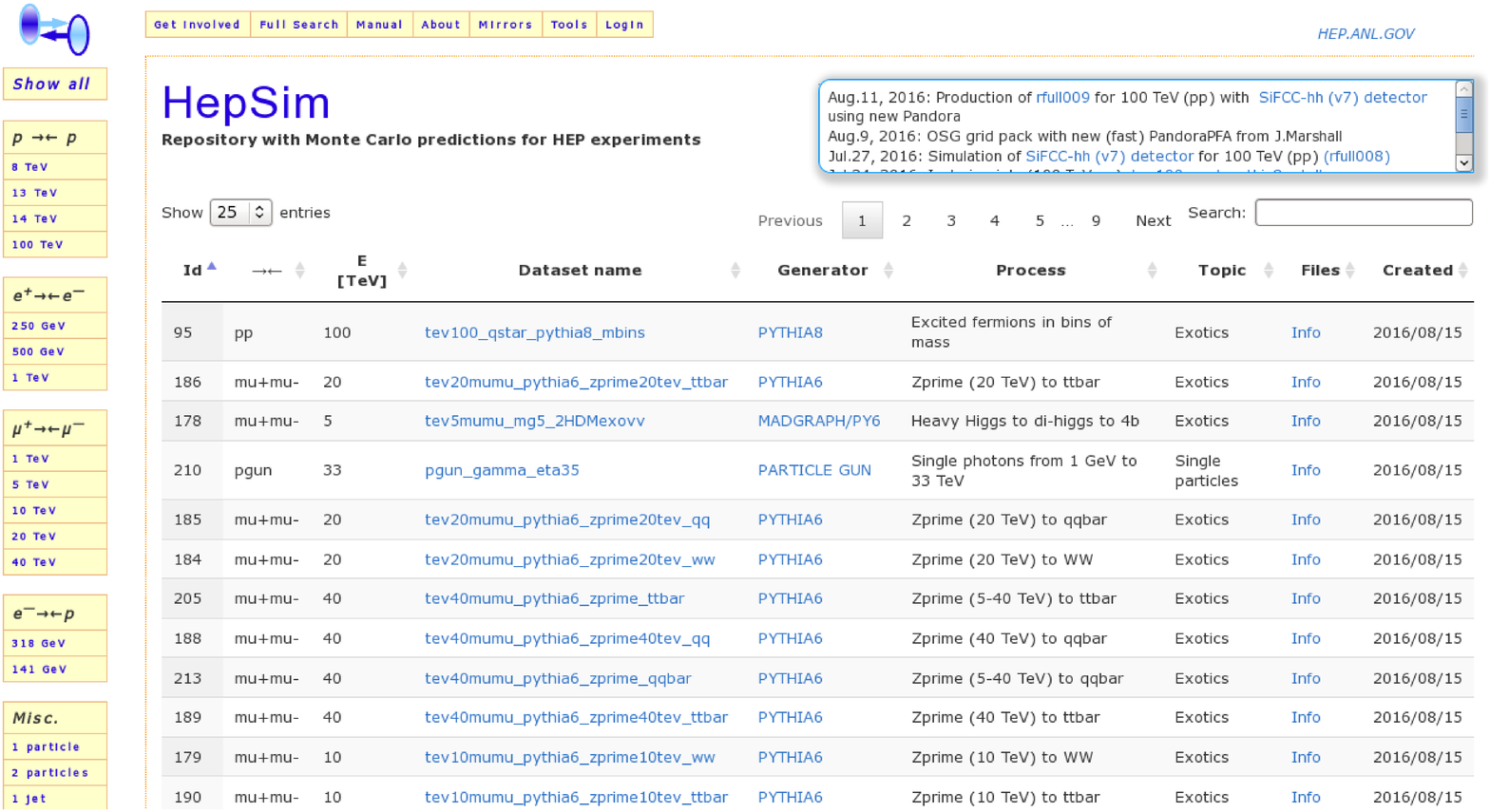}
\caption{Illustration of the front-end of HepSim where a user can select the needed process.}
\label{fig:hepsim}
\end{figure}
 
The user front-end of the HepSim database is shown in Figure~\ref{fig:hepsim}. 
The web interface of HepSim allows searching for the needed processes and datasets.
The files can be downloaded using the ``hs-tools'' Java package as described in \cite{Chekanov:2015cca}.
In addition to the user-friendly set of tools for searching and downloading files, the package can be used 
for a quick processing of EVGEN events in order to create the 
needed kinematic distributions and cross sections for physics studies. 
For example, the tool includes histogram classes, plot canvases for 2D/3D plots, and particle classes and jet algorithms \cite{Chekanov:2015cca}.
Data analysis can be performed using the Jas4pp program \cite{jas4pp}, which 
originates from the Jas3 program developed at SLAC for the ILC.
This program supports
data analysis of EVGEN events, as well as analysis of the 
data  after full simulations to be described later. This package also provides the Wired4 event display.
The common Java API shared by the Jas4pp and ``hs-tools'' is
described in \cite{Chekanov:Book2}.
To perform a complete data analysis, the needed C++ packages are also available from the ProMC package \cite{2013arXiv1311.1229C}.

\subsection{EVGEN events}
Created EVGEN events 
are kept in the archive file format called ProMC \cite{2013arXiv1311.1229C}. 
This binary format was designed
to keep unweighted events from parton-shower event generators and  weighted events from  
next-to-leading order QCD calculations. In addition, the original log files created during the generation step are included.
To keep file sizes small, a variable-length integer encoding scheme (``varints'') is used. 

Open access to ProMC files with EVGEN events is the central part of the HepSim design, since
further fast and full event simulations are performed using computer resources from multiple locations. 
The compact files designed for web streaming, together with the http protocol optimized 
for handling many (relatively small) files, is one of the distinct features of HepSim compared to other production systems.

\subsection{Full Geant4 simulations}
A number of detector descriptions have been created  for HepSim using the 
``silicon detector'' (SiD) detector designed for the ILC project.  
This detector combines excellent silicon tracking with high-granular electromagnetic and hadronic calorimeters which 
are required  for
identification of separate particles. 

\begin{figure}
\centering
\includegraphics[width=.7\textwidth]{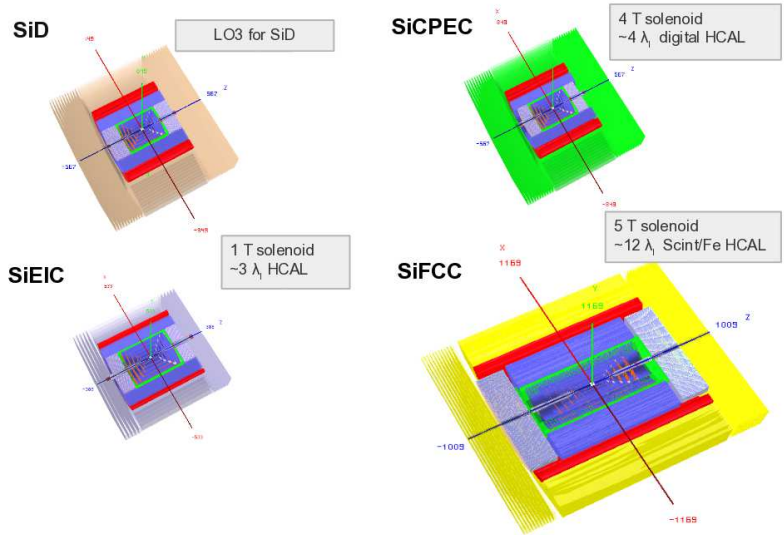}
\caption{
This graphic shows cross-sections of the four detectors available from HepSim. 
They are all designed using the ``all-silicon'' concept inspired by the SiD detector.
The red color shows the solenoid which encloses hadronic, electromagnetic and outer and inner tracker. The muon detector is
located behind the solenoid. This image retains the relative sizes of the shown detectors. } 
\label{fig:detectors}
\end{figure}

A family of ``all-silicon`` detectors was created using the ``compact.xml'' geometry files in the XML format.
Such geometry files can be converted to the Geometry Description Markup Language (GDML) format for easy 
viewing and debugging using ROOT.
Figure~\ref{fig:detectors} shows the cross sections of four detectors available from the HepSim GDML repository.
Some of these detectors, such as SiFCC (a ``performance'' detector for a 100 TeV collider), have several production versions.

The response of the detectors to physics processes
is simulated using 
the Simulator for the Linear Collider (SLIC) package \cite{Graf:2006ei} 
developed for the ILC project.
The main strength of this software lies in the fact that
detector geometries can easily be changed using the XML files. 
The final output of the reconstruction step is 
the particle-flow objects created by the 
Pandora Particle Flow algorithm  \cite{Charles:2009ta,Marshall:2013bda}.
In addition, complete  information on reconstructed tracks, calorimeter clusters and calorimeter hits is preserved for data analysis. The reconstructed events are kept in the original SLIC format, called LCIO \cite{lcio}. 

Currently, less than 0.01\% of EVGEN events hosted by HepSim were simulated and reconstructed using HepSim detector
designs. These  simulations are typically required to verify detector performances and detector response to 
physics processes. 
The processing step ProMC$\to$LCIO  that  creates events with fully reconstructed tracks, 
calorimeter clusters and particle flow objects
can take up to 4 CPU*hours per event for 100 TeV $pp$ collisions (without pileup) in the SiFCC detector 
which features the high-granularity calorimeters and silicon trackers.
A memory usage of up to 16 GB is required. 
We do not expect a significant increase of the number of simulated and reconstructed events in the near
future due to the lack of general-purpose computing resources. 
 
Data analysis of fully simulated and reconstructed events  can be done with the platform-independent Jas4pp program 
using the Java or Python programming languages.   
There is also a dedicated C++ package to be compiled on the Linux platform.  

\subsection{Fast simulations}
The fast detector simulation  uses the DELPHES framework \cite{deFavereau:2013fsa} that incorporates a
tracking system, magnetic field, calorimeters, and a muon system. 
DELPHES simulates the calorimeter system by summing together 
cells to form towers\footnote{Towers are regions defined in  pseudorapidity $\eta$ and
azimuthal angle $\phi$.}.
Currently, three types of detectors are supported: a CMS-like detector, an ATLAS-like detector, and the 
detector developed during the US long-term planning study 
of the future program of particle physics (Snowmass 2013) \cite{Anderson:2013kxz}.
The output from the fast simulation is kept in the ROOT files. About 1\% of the EVGEN files were
processed with the DELPHES program. The fast simulation is more than a factor $10^5$ faster compared to the 
full SLIC simulation described above, but DELPHES simulation often lacks important information needed 
for understanding impact of detector performance on physics processes.

\section{Conclusions}
To increase the pace of scientific discoveries in high-energy particle physics and 
to foster designs of new HEP experiments, 
HepSim datasets  can freely be 
downloaded for physics and detector-performance publications without any restrictions. 
The HepSim Monte Carlo data repository has been readily accessible to the public since 2014.
In addition
to the data files, the HepSim online manual contains code snippets, which can easily be 
incorporated into data-analysis projects.  
Recently, there were several studies based on HepSim data:  CEPC studies \cite{marcel} using the ``all-silicon'' concept based on the SiD detector are 
exclusively based on HepSim samples. 
Single-particle samples for the ATLAS tracking upgrade for the HL-LHC experiment were created using the HepSim tools.
Several studies based on full detector simulations 
included in HepSim 
were presented during this conference \cite{study1,study2,study3}. The repository was also used to initiate 
the creation of Monte Carlo samples for several physics papers by the ATLAS collaboration. 
Currently, HepSim is an indispensable tool for FCC-hh physics and detector performance studies
since the unprecedented energy regime imposes new requirements on detector technologies which can be studied using 
detailed Geant4 simulations.

\section*{Acknowledgments}
The author would like to thank many people who provided software codes and Monte Carlo samples for inclusion  into the HepSim repository.
Their names are listed on the HepSim web page.  
Argonne National Laboratory's work was supported by the U.S. Department of Energy, Office of Science under contract DE-AC02-06CH11357.



\begin{thebibliography}{99}


\bibitem{Chekanov:2014fga}
S.~Chekanov, \emph{HepSim: a repository with predictions for high-energy physics
  experiments}, \emph{Advances in High Energy Physics}, {\bf 2015} (2015) 136093, available
  as \url{http://atlaswww.hep.anl.gov/hepsim/}.

\bibitem{Chekanov:2015cca}
S.~V. Chekanov, I.~Pogrebnyak, D.~Wilbern, \emph {Cross-platform validation and
  analysis environment for particle physics}, (2016) (submitted to Comp. Phys. Comm.), 
\href{http://arxiv.org/abs/1510.06638}{\path{arXiv:1510.06638}}. 

\bibitem{jas4pp}
\emph {Jas4pp. Java Analysis Studio for Particle Physics},
  \url{https://atlaswww.hep.anl.gov/asc/jas4pp/} (2016).

\bibitem{Chekanov:Book2}
S.~V.~Chekanov, \emph{Numeric Computation and Statistical Data Analysis on the Java
  Platform}, Springer, London, 2016, \emph{ISBN 978-3-319-28531-3}.

\bibitem{2013arXiv1311.1229C}
S.~V. Chekanov, E.~May, K.~Strand, P.~Van~Gemmeren, \emph{ProMC: Input-output data
  format for HEP applications using varint encoding}, \emph{Comp.~Phys.~Commun.} {\bf 185} 
  (2013) 2629.

\bibitem{Graf:2006ei}
N.~Graf and J.~McCormick, \emph{Simulator for the linear collider (SLIC): A tool for
  ILC detector simulations}, \emph{AIP Conf. Proc.} {\bf 867} (2006) 503.


\bibitem{Charles:2009ta}
M.~J. Charles, \emph{PFA Performance for SiD}, in: \emph{Linear colliders. Proceedings,
  International Linear Collider Workshop, LCWS08, and International Linear
  Collider Meeting, ILC08, Chicago, USA, Novermber 16-20, 2008 }, 2009.
\newblock \href {http://arxiv.org/abs/0901.4670} {\path{arXiv:0901.4670}}.

\bibitem{Marshall:2013bda}
J.~S. Marshall and M.~A. Thomson, \emph{Pandora Particle Flow Algorithm}, in:
  \emph{Proceedings, International Conference on Calorimetry for the High Energy
  Frontier (CHEF 2013)}, 2013, pp. 305.

\bibitem{deFavereau:2013fsa}
J.~de~Favereau, et~al., \emph{DELPHES 3, A modular framework for fast simulation of
  a generic collider experiment}, \emph{JHEP} {\bf 1402} (2014) 057.

\bibitem{Anderson:2013kxz}
J.~Anderson, et~al.,
\emph{Snow
mass Energy Frontier Simulations}, in: \emph{Proceedings, Community Summer Study
  2013: Snowmass on the Mississippi (CSS2013)}, Minneapolis, MN, USA, July
  29-August 6, 2013, 
  \newblock \href {http://arxiv.org/abs/1309.1057} {\path{arXiv:1309.1057}}.

\bibitem{marcel}
S.~V. Chekanov and M. Demarteau, \emph{Conceptual Design Studies for a CEPC Detector}, 
\newblock \href {http://arxiv.org/abs/arXiv:1604.01994} {\path{arXiv:1604.01994}}.
(2016). A white paper contributed to the IAS Program on High Energy Physics (4-29 Jan, 2016).

\bibitem{lcio}
F.~Gaede et~al., \emph{LCIO: A persistency framework for linear collider simulation studies},
Talk from the 2003 Computing in High Energy and Nuclear Physics (CHEP03), La Jolla, Ca, USA, March 2003,
\newblock \href {https://arxiv.org/abs/physics/0306114} {\path{arXiv:physics/0306114}}. 
 

\bibitem{study1}
Y.~Shin-Shan, et~al., \emph{Study Of Boosted W-Jets And Higgs-Jets With the SiFCC},
A talk given at the ICHEP16 conference. 

\bibitem{study2}
S.~Sen, et~al., \emph{Tau reconstruction for a FCC experiment}, ICHEP16 poster.

\bibitem{study3}
N.~V. Tran, et~al., \emph{High-granularity calorimeter for a FCC detector}, ICHEP16 poster.

\end{thebibliography}
\end{document}